\documentclass[12pt]{article}

\usepackage{geometry}
\usepackage{amsmath,amssymb}
\usepackage{graphicx}
\usepackage{booktabs}
\usepackage{tabulary} 
\usepackage{array}    
\usepackage{natbib}   
\usepackage{hyperref}
\usepackage{multirow}
\usepackage[table]{xcolor} 
\definecolor{MyLightBlue}{RGB}{220, 240, 255}
\definecolor{MySoftYellow}{HTML}{FFF2CC}
\definecolor{MySoftGreen}{RGB}{226, 239, 218}
\usepackage[english]{babel}
\usepackage[autostyle=true]{csquotes}
\usepackage[T1]{fontenc}
\usepackage{lmodern}
\usepackage{tcolorbox}
\tcbuselibrary{listings,breakable,skins}
\newtcolorbox{promptbox}{
  breakable,
  enhanced,
  colback=gray!5,
  colframe=black,
  title=LLM Prompt,
  fonttitle=\bfseries,
  verbatim,
  left=1mm,
  right=1mm,
  top=1mm,
  bottom=1mm
}
\usepackage{pifont} 
\usepackage{xcolor}

\title{Decision Models as Scalable Economic Surrogates that Capture Human Variation}

\author{
  \small
  \begin{tabular}{p{\textwidth}}
    \centering
    Maxime Saxena$^{1}$, Fabio Caccioli$^{1,2}$, \mbox{and R. Maria del Rio-Chanona$^{3, 1,\dagger}$} \\[1em]
    \small $^{1}$Department of Computer Science, University College London, UK \\
    \small $^{2}$ Systemic Risk 
    Centre, London School of Economics and Political Science, UK\\
    \small $^{3}$Google DeepMind \\
  \end{tabular}
}

\date{\today \vspace{-2em}}

\begin{document}
\maketitle

\renewcommand{\thefootnote}{$\dagger$} 
\footnotetext{Correspondence: m.delriochanona@ucl.ac.uk}
\renewcommand{\thefootnote}{\arabic{footnote}} 

\begin{abstract}
Decision Models (DMs), a new class of AI models, give fast responses and return a probability distribution over a fixed set of options. These properties make them candidate human surrogates in economic simulations and surveys, a role increasingly filled by Large Language Models (LLMs), which are costly at scale and under-represent the variation in human behaviour. We present a first evaluation of DMs as human surrogates in economics, comparing 10 DMs with 8 LLMs in three economic settings: a household spending survey on consumption, several classic economic games and a labour-market field experiment. Frontier DMs, those that score highest on general decision-making benchmarks, reproduce the variation in human responses more closely than LLMs in all three settings. They capture the modal human response as well as LLMs in the spending survey, but less well in the economic games. Frontier DMs are also roughly ten times faster than the fastest LLM and can be over a hundred times cheaper. Together, these results suggest that DMs capture the heterogeneity of human behaviour better than LLMs and make large-scale economic simulations more feasible.
\end{abstract}

\section{Introduction}
\label{sec:intro}

AI agents are increasingly used as human surrogates in surveys or agent-based simulations, with recent applications spanning political preferences \citep{argyle2023out}, consumer choices \citep{maier2025llms} and strategic economic games \citep{horton2023large,mei2024turing}. AI agents address the core challenge in socio-economic modelling of balancing agent adaptability with behavioural realism. Traditional agent-based simulations rely on fixed, hand-written behavioural rules \citep{farmer2009economy}. Because people adapt when policy changes, these rules may not generalise to new policies, a problem known as the Lucas critique \citep{lucas1976econometric}. To address this critique, economic models have used perfectly optimising rational expectations agents, an assumption that does not align with real human behaviour \citep{conlisk1996bounded, kahneman2013prospect, thaler2016behavioral}. While surrogates such as LLM-agents are imperfect approximations of humans \citep{gao2025take}, they can capture many human behavioural biases while adapting dynamically to policy changes \citep{horton2023large}. They therefore provide a much-needed alternative to the unrealistic behavioural assumptions of the status quo, enabling simulations that can generate more reliable answers to policy questions.

However, using LLMs as human surrogates faces two major limitations. First, while LLMs often successfully replicate the modal human response, they fail to capture the full distribution of behaviour, leading to significantly less heterogeneity than is found in human populations \citep{bisbee2024synthetic, wang2025large, wu2026position}. This compromises economic modelling for policy, as human heterogeneity is a driver of both macroeconomic outcomes \citep{kirman1992whom} and collective social dynamics \citep{granovetter1978threshold, wu2026position}. Although persona prompting and demographic conditioning can recover some of this lost variation \citep{vezhnevets2023generative}, they require extensive sampling, which leads to the second challenge: operational latency and computational cost. Generating autoregressive text for multiple, sequential decisions across large-scale populations is slow and computationally expensive, forcing researchers to trade off the size of their simulated economies against agent diversity.

Decision Models (DMs), a rapidly growing class of AI models \citep{duan2026general, ling2026jev, almeida2026jev}, are in a good position to address these challenges. They are fine-tuned to take as input natural language questions and scenarios and make decisions between provided options. Crucially, rather than generating free-form text, DMs are designed to directly assign probabilities to the available choices.  They are also faster and cheaper than LLMs, potentially allowing simulations with orders of magnitude more agents. This makes them a promising candidate for representing human behaviour as they are adaptable, provide structured choice probabilities, are fast and can be sampled cheaply to generate variation in simulated decisions.

In this paper, we test whether DMs are a viable additional tool for representing humans in economic modelling.  We choose our experiments so that we explore a range of economic settings (consumption, games and labour) and question types (binary choices vs multiple choices and qualitative vs quantitative).  The human data we compare to varies from surveys to economic experiments and realistic field experiments. We consider survey answers on changes in consumption in response to a future earnings shock \citep{frbny_sce_spending}, a variety of classic economic games \citep{mei2024turing} and a field experiment on labour preferences \citep{mas2017valuing}. By replicating these experiments and surveys with DMs and LLMs, we assess how well DMs are able to act as human surrogates for economic research compared to LLMs. 

Our results show that \enquote{frontier} DMs, those that lead on standard decision-making benchmarks, capture the variation in human behaviour better than LLMs, and do so faster and more cheaply. In all three settings, frontier DMs are closer to the human distribution than LLMs on our distributional measures. In the consumption survey, they also capture the modal human response as well as LLMs, although on the economic games, LLMs represent the modal human response better. In the labour experiment, frontier DMs reproduce the gradual shift in human choices as a wage gap changes, whereas LLMs switch abruptly from one job to the other. Taken together, we find that frontier DMs can be an effective and scalable tool for modelling human behaviour, especially where capturing the variation in responses matters.

\subsection{Related work}

Our work fits into a broader agenda of integrating AI models into economic research. Reinforcement Learning has been used in macroeconomics \citep{gabriele2025heterogeneous} and firm competition environments \citep{calvano2020artificial}. A growing body of work has also studied LLMs, both to understand their behaviour and to assess whether they are viable human surrogates, with encouraging results across a range of single-shot and multi-turn games \citep{horton2023large, mei2024turing, manning2026general}. LLMs have also been used within agent-based models (ABMs), in diverse settings such as macroeconomics \citep{li2024econagent, chen2025hierarchical, feng2025simcity}, status signalling \citep{cross2026generative} and trading markets \citep{henning2025llm, del2025can, saxena2026machine}, with platforms such as Concordia allowing such simulations to be built quickly and flexibly \citep{vezhnevets2023generative}.

However, the use of LLMs as human surrogates remains contested: LLMs tend to capture only a narrow, modal part of the human distribution \citep{bisbee2024synthetic, wang2025large, wu2026position}, and the outcomes of LLM-based ABMs can be sensitive to prompting \citep{ye2026stop}. Several works have therefore aimed to make LLMs better represent human distributions. For example, conditioning LLMs on personas or backstories can widen the part of the human distribution they capture \citep{argyle2023out, park2024generative}, although this approach can be brittle to the exact wording of the prompt \citep{lutz2025prompt}. 

Work on LLM-agent simulations has also been limited in scale, typically to populations of a few hundred or a thousand agents. However, complex systems such as the economy of a city or country have millions of agents. Some progress has been made on scaling LLM-based ABMs with the number of agents going over the million mark, although this remains the exception, not the rule. For example, \citet{chopra2024limits} used archetypes to achieve scale, where a single query determines the behaviour of a group of agents. Alternatively, some ABMs increase scale by not having every agent act at every time-step \citep{yang2024oasis}. However, these approaches still involve a trade-off between a more faithful simulation requiring more LLM calls and the time required for the simulation. 

These limitations should nonetheless be weighed against the alternatives. Economics lacks a widely accepted model of human behaviour with the \enquote{full information rational expectations} and perfectly optimising agents assumptions \footnote{Agents have full information, model-consistent future expectations and successfully optimise their utility.} still prevalent. This is despite substantial evidence from behavioural economics suggesting these assumptions are a poor model for humans \citep{conlisk1996bounded, dellavigna2009psychology, coibion2018formation}. Although LLMs are not a perfect model, they do offer a sensible alternative. 

DMs may also be a viable alternative surrogate for humans. We are, to the best of our knowledge, the first to research using DMs as human surrogates for economic modelling. The closest work to ours is \citet{huang2026jevolution}, who used Jev to make decisions inside a predator-prey simulation. DMs have also been considered for use in social sciences by \citet{ibrahim2026evaluating}, who found that DMs can be useful as a first step in annotation pipelines. On DM evaluation, \citet{asseri2026calibrated} used the Values Survey Module 2013 to find that Jev, one of the most popular DMs, defaults to American values.  However, we expect the body of work on DMs to expand quickly as they become more entrenched in the AI ecosystem.

\section{Methods}
\label{sec:methods}

\subsection{Experiments}

We start with a survey of US households run every 4 months by the Federal Reserve Bank of New York. We use questions from the Household Spending Survey, which ask how an unexpected 10\% increase and 10\% decrease in future earnings would impact household consumption \citep{frbny_sce_spending}. For each question, there are 7 possible answers, such as \enquote{Spend some and save some} or \enquote{Cut spending by some and cut savings by some}. Further details can be found in Appendix \ref{app:experiments}. We randomly select 1000 households from the 2020-2024 data in this experiment.

Economic literature suggests that stated preferences in surveys do not always accurately reflect revealed preferences, what people actually do \citep{cummings1995homegrown, list2001experimental, murphy2005meta}.  Therefore we also consider two revealed preference datasets, where respondents take actions and their preference is then inferred from these actions. We start with economic games, where a short scenario is presented to a participant and they must choose their action. For example, in the \enquote{Bomb Risk Game}, the participant is given 100 boxes. They win a dollar value proportional to the number of these 100 boxes they decide to open but hidden in one of these boxes is a \enquote{bomb}; if the bomb is in a box they choose, they receive nothing. Participants then choose how many boxes to open as a number from 0-100.

We use five economic games found in \citet{mei2024turing}. Two of the five games have two roles, so there are a total of 7 \enquote{positions} to test from these games. We use the following games which test the attributes in brackets: Dictator Game (altruism), Ultimatum Game (two positions testing fairness and spite), Trust Game (two positions testing trust, fairness, altruism and reciprocity), Bomb Risk Game (risk aversion) and Public Goods Game (free-riding, altruism and cooperation). Details of these games are in Appendix \ref{app:experiments}. In all 7 positions, participants choose between numbers 0-100 or 0-150 depending on the position.

For the third dataset, we use the natural field experiment of \citet{mas2017valuing}, which is the closest setting to real human choices we consider. In this experiment, participants applied to a call centre job and as part of the application were asked to choose between two jobs. Both jobs involved identical work but had differing working arrangements: one job was always a standard 9-5, 40 hour week job done in person at an office, while the other had an alternative working arrangement.  The five alternatives were: flexible schedule (changing times and days of work), flexible number of hours (up to 40 hours a week), the option to work from home, a combined flexible option with all of these, and an arrangement where the employer chose hours and schedule one week in advance. As applicants saw these two jobs, the wages of the two jobs were also varied, so for example the standard job might be paid at \$19 per hour, while having the option to work from home paid \$16.5 per hour. 

Once these choices are made, we fit a logistic curve to the data to give the proportion of respondents who chose the alternative schedule as the wage gap between the alternative and standard job varied. This gives \enquote{Willingness to Pay} (WTP) curves which are effectively cumulative distributions describing the proportion of respondents who would choose the alternative if the alternative paid $x$ dollars less (more) than the standard job. However, \citet{mas2017valuing} found that a significant proportion (25-30\%) of human participants were not paying attention and choosing at random, estimated by the number of people who chose the less flexible job even when it paid \$5 less. All WTP calculations therefore first estimate and remove this inattention before fitting the curve. Details are in Appendix \ref{app:experiments}.

\subsection{Running experiments on DMs and LLMs}

DMs, also known as \enquote{System 1} models in reference to Kahneman's intuitive decision making \citep{kahneman2011thinking}, are fast, cheap models specialised in making decisions. They take a natural language description of a situation, a question and a set of options and return a probability for each option supplied in one forward pass. These probabilities form a distribution over the supplied options by construction. The types of decisions they can make are multiple choice (\enquote{choice} format), a level on an ordered scale (\enquote{score} format) or a binary yes/no (\enquote{noul} format). Technical summaries are thin but the interested reader is referred to \citet{zhang2026anyjev,geuter2026releasing, hume2026jevarchitecture}.

We consider 10 DMs across a range of providers. These models span capability levels in standard Decision benchmarks, and we use the Artificial Analysis Decision Index to define a group of \enquote{frontier} DMs, consisting of Jev, D1 Ultra, Mammoth and Perplexity \citep{artificialanalysis2026decisionindex, kirkby2026mammoth}. These four models are the four highest scoring DMs of those we test on the Artificial Analysis Decision Index. Model details and a discussion of DM benchmarks are in Appendix \ref{app:models}. 

For experiments in \citet{mei2024turing} with numerical answers (i.e. pick a number between 0-100), we define 10 evenly spaced bins. We test DM responses as both a choice over the middle of the bins (5,15,25 etc) or a score over the bins (\enquote{0-9}, \enquote{10-19} etc). The difference between these DM response modes is that scores remain ordered and the DM knows that bin 2 is after bin 1. In a choice, the ordering of the bins does not matter. For all other experiments, we use a simple choice as in the original experiments, either between two options for \citet{mas2017valuing} or 7 options for \citet{frbny_sce_spending}.

For LLMs, we use \citet{mei2024turing}'s method of eliciting answers and consider a total of 8 LLMs, of which for two we also look at varying their reasoning parameters. These models span a range of sizes and include models from the GPT, Claude, Gemini, Gemma, GLM and Qwen families. We use recommended or default settings for LLMs. Details of the models and their answer formats are in Appendix \ref{app:models}. 

Since neither DMs nor LLMs give deterministic outputs, we average results over repeats of the experiments. The consumption survey has demographic data attached to each answer and so we create 1000 personas using data on age, sex, race, US state, education, marriage, whether they are employed and income bracket. We prompt models with these personas as well as the question. Since we consider 1000 separate personas, we do not do repeats for each persona. For the economic games, LLMs and humans give continuous numerical answers while DMs can only answer with a choice between 10 bins. We therefore bin the LLM and human answers into these 10 bins for comparison. We do 50 repeats for each game and every DM and LLM. DM distributions are then averaged over the repeats for each game while LLM distributions are simply the binned distribution of the 50 LLM answers per game. In the labour choice experiment, every model makes each choice 6 times. We consider a high and a low wage regime and also ask questions with the binary choices in both orders. We consider 29 different wage gaps, meaning that each of the 5 treatments has 696 responses in total per model.

\subsection{Measuring agreement with human results}

We use a Turing Test, following \citet{mei2024turing}, to measure how well a model captures the modal human answers. This test asks: what is the probability that if we pick a random model data point and a random human data point, the model data point will be more typical under the human distribution? Unlike in \citet{mei2024turing}, we calculate the exact expected win rate of the random model data point rather than randomly sampling and recording the wins. The Turing Test rewards picking the modal human outcome, since a model which only ever picks the modal human outcome can never be less likely than another data point under the human distribution. However, a high win rate does not mean a good distributional fit; a model which only picks the modal human answer ignores the rest of the human distribution but scores highly on the Turing Test. This is why some models seem to \enquote{beat} humans in being more human-like. 

For one question (or game and role), let the answer options be $k = 1, \dots, K$. Let $p_k$ be the share of humans who chose option $k$, and $q_k$ the share of a model's answers on option $k$. Draw one model answer $a \sim q$ and, independently, one human answer $h \sim p$. The Turing Test win-rate is then:
\begin{align}
  W &= \Pr(p_a > p_h) = \sum_{k=1}^{K} \sum_{l=1}^{K} q_k \, p_l \, \mathbf{1}\{p_k > p_l\}.
\end{align}
We compare the DM and LLM Turing Test scores with the human \enquote{self-play} score, where $q$ is set equal to $p$.

We use two measures of distributional fit throughout the paper. Where options are naturally ordered, as in the economic games and the labour choices, we use the Wasserstein-1 distance, also called the \enquote{earth-movers} distance. Where appropriate, we check these results also hold with Kullback-Leibler (KL) divergence. A lower Wasserstein-1 distance denotes a closer agreement with the target human distribution, and we normalise over the width of the range of answers to make all answers comparable.  By definition, the Wasserstein-1 distance of the human distribution from itself is 0. The Wasserstein-1 distance between two probability distribution functions, $u,v$, with Cumulative Distribution Functions, $U,V$, is defined as:

\begin{align}
W_1(u, v) = \int_{-\infty}^{\infty} \left| U(x) - V(x) \right| \, \mathrm{d}x .
\end{align}

For comparison of two probability distributions where the choices have no natural distance measure between them (as in \citet{frbny_sce_spending}), we use KL divergence. As with the Wasserstein-1 distance, closer to 0 is a better fit and humans have a KL divergence of 0 from themselves by definition. The KL divergence between two probability distribution functions, $u,v$, is defined as:

\begin{align}
D_{\mathrm{KL}}(u \,\|\, v) = \sum_{i=1}^{n} u_i \log \frac{u_i}{v_i}.
\end{align}
To account for the fact that the KL divergence can go to infinity if any option has zero samples, we mix all distributions with 1\% of a uniform distribution.

\section{Results}
\label{sec:results}

\subsection{Consumer spending}
\label{subsec:consumer}

\begin{figure}[htbp]
    \centering
    \includegraphics[width=\linewidth]{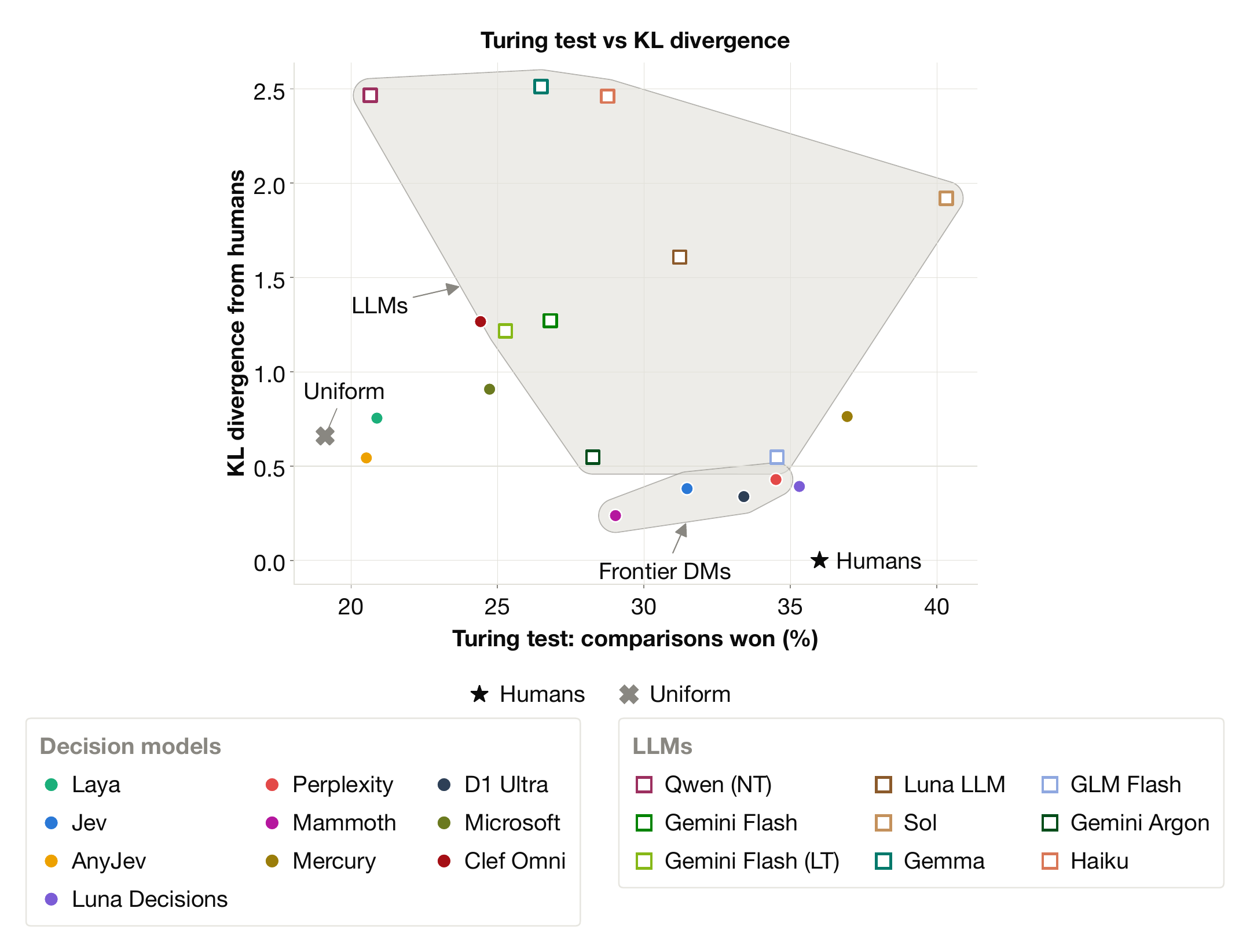}
    \caption{\textbf{Frontier DMs better represent humans than LLMs in two consumer spending survey questions.} KL divergence is compared to the Turing Test. DMs are solid circles while LLMs are hollow squares.  Scores are averaged over the two consumer spending survey questions. Qwen (NT) is Qwen with no thinking, and Gemini Flash (LT) is Gemini Flash with low thinking.}
    \label{fig:sce}
\end{figure}

Starting with the survey questions in \citet{frbny_sce_spending}, Figure \ref{fig:sce} shows that frontier DMs capture human responses more closely than both LLMs and other DMs. On the mean KL divergence across the two questions, every frontier DM is closer to humans than every LLM: all 36 pairwise comparisons (4 DMs, 9 LLMs) are significant at the 1\% level under a two-sided bootstrap test with Holm correction.

The Turing test is more mixed: using the same significance tests as above, frontier DMs have significantly higher win rates in 6 tests, significantly lower in 7 and no significant difference in 23 tests compared to the 9 LLMs. On this experiment, frontier DMs and LLMs do not differ systematically in capturing the modal human, as measured by the Turing Test. Details of all significance test results, including those showing that frontier DMs are better than the Uniform baseline on both measures, are in Appendix \ref{app:consumer}. The Turing Test has fewer significant tests because its win rates are far more sensitive to the resampling from the bootstrap procedure than the KL divergence metric, meaning that the errors on the metrics are wider. This sensitivity is a good reason to include distributional measures like KL divergence, which are less binary than a win or no-win outcome and therefore vary more smoothly with resampling than the Turing test. 

We consider a range of robustness checks, looking at minor changes to the wording of one question, asking the models to \enquote{act like a human} and exploring the impact of the system prompt for LLMs. We run these checks on the four frontier DMs (Mammoth, Jev, D1 Ultra and Perplexity) and three LLMs (Gemini Argon, GLM Flash and Sol) and find that our results are robust to these changes; full details are in Appendix \ref{app:consumer}.

Turning to non-frontier DMs, Laya and AnyJev perform close to the Uniform benchmark on both Turing test and KL divergence. Other DMs such as Mercury or Luna Decisions, which are behind the DM frontier according to Artificial Analysis, beat the uniform baseline on the Turing test but only Luna Decisions has a lower KL divergence than the Uniform baseline. For this task, Luna Decisions is as good at capturing the human behaviour as the frontier DMs. 

Most LLMs do not fit the human distribution well on this task. All LLMs except GLM Flash and Gemini Argon have a higher KL divergence than the Uniform benchmark (0.66 vs 1.22-2.47), suggesting that they tend to focus their answers more consistently on a narrower range of answers. However, GLM Flash and Argon both fit the human mode and distribution well, although they are still statistically significantly worse than frontier DMs on the KL divergence metric. On this task, frontier DMs (and Luna Decision) are therefore the most faithful human surrogates we test, matching both the spread of human answers and the typical answer.

\subsection{Economic games}
\label{subsec:economicgames}

\begin{figure}[htbp]
    \centering
    \includegraphics[width=\linewidth]{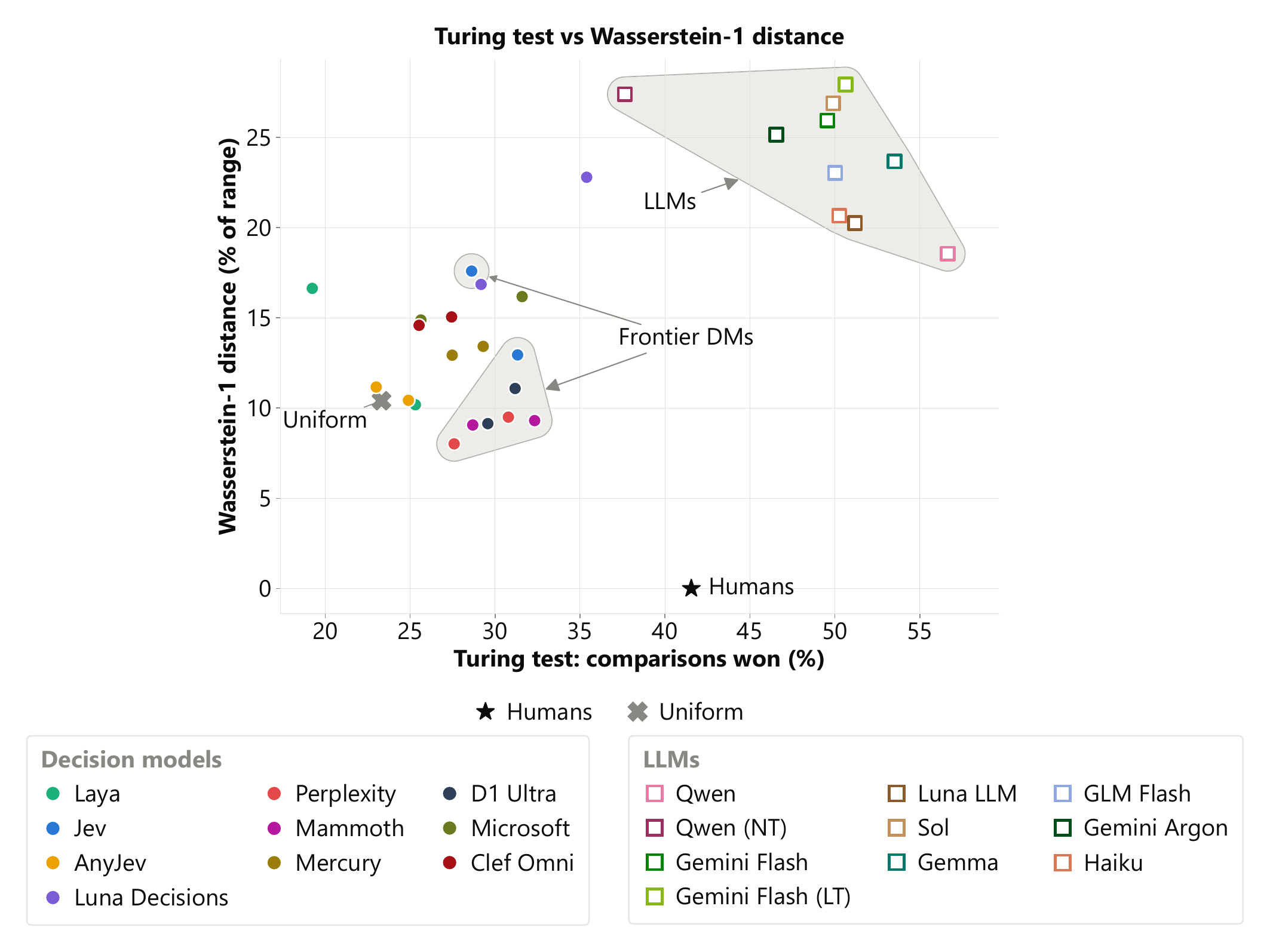}
    \caption{\textbf{Frontier DMs capture the full human distribution better.} Mean results from the 7 positions in \citet{mei2024turing}. Shows the Wasserstein-1 distance and the Turing Test results against the Human results and a Uniform benchmark. DMs are circles and LLMs are hollow squares. The frontier DMs are: Jev, D1 Ultra, Mammoth and Perplexity. Qwen (NT) is Qwen with no thinking, and Gemini Flash (LT) is Gemini Flash with low thinking.}
    \label{fig:mei}
\end{figure}

Figure \ref{fig:mei} shows that frontier DMs capture the distribution of human results better than LLMs according to the mean Wasserstein-1 metric. Using a bootstrapping procedure and a two-sided test between all pairs of frontier DMs and LLMs (80 tests, with 4 DMs, 2 answer methods per DM and 10 LLMs), 79 tests show frontier DMs with statistically significantly lower mean Wasserstein-1 distance to the human distribution than LLMs (1\% level with Holm correction, all 7 positions equal weights), and only 1 test shows no significant difference. As a robustness check using KL divergence rather than Wasserstein-1, all 80 tests show frontier DMs with statistically significantly lower KL divergence than LLMs; this is because the KL divergence penalises bins where a model has zero answers strongly, therefore skewing the results in favour of DMs. Details of these tests are in Appendix \ref{app:economicgames}.

However, LLMs have higher Turing Test win rates in these games. Using the same bootstrap procedure as above, all LLMs have statistically significantly higher Turing test win rates than the frontier DMs (full results in Appendix \ref{app:economicgames}). This shows that LLMs are better at concentrating near the modal human responses on average than DMs for these games. However, the fact that several LLMs have higher Turing Test win rates than humans also supports the results with the Wasserstein-1 distance; LLMs tend to overly concentrate near the modal human answer while DMs capture a wider spread of the human distribution.

We find that although using the score vs choice answer method does change DM results slightly, it does not do so in a consistent way across differing DMs.

Just as with LLMs, not all DMs are equal. The frontier DMs (Jev, D1 Ultra, Mammoth and Perplexity) generally have lower Wasserstein-1 distance and higher Turing test metrics than other DMs. All DMs which have statistically significantly lower Wasserstein-1 distance than the Uniform distribution are frontier DMs and frontier DMs also have a higher Turing test win rate than the baseline. As in Section \ref{subsec:consumer}, Laya and AnyJev are close to the Uniform benchmark while the remaining DMs are somewhere between the Uniform benchmark and the frontier models.

The uniform baseline is a strong baseline on the Wasserstein-1 measure for this data, partially because the human results at times are also close to uniform. A good representation of human behaviour needs the combination of capturing the modal human answers as well as its broader distribution. LLMs focus on the modal response and in some senses, DMs therefore provide a middle ground between the uniform baseline and the modal-response focus of LLMs. 

\subsection{Labour choices}
\label{subsec:labour}

\begin{figure}[htbp]
    \centering
    \includegraphics[width=\linewidth]{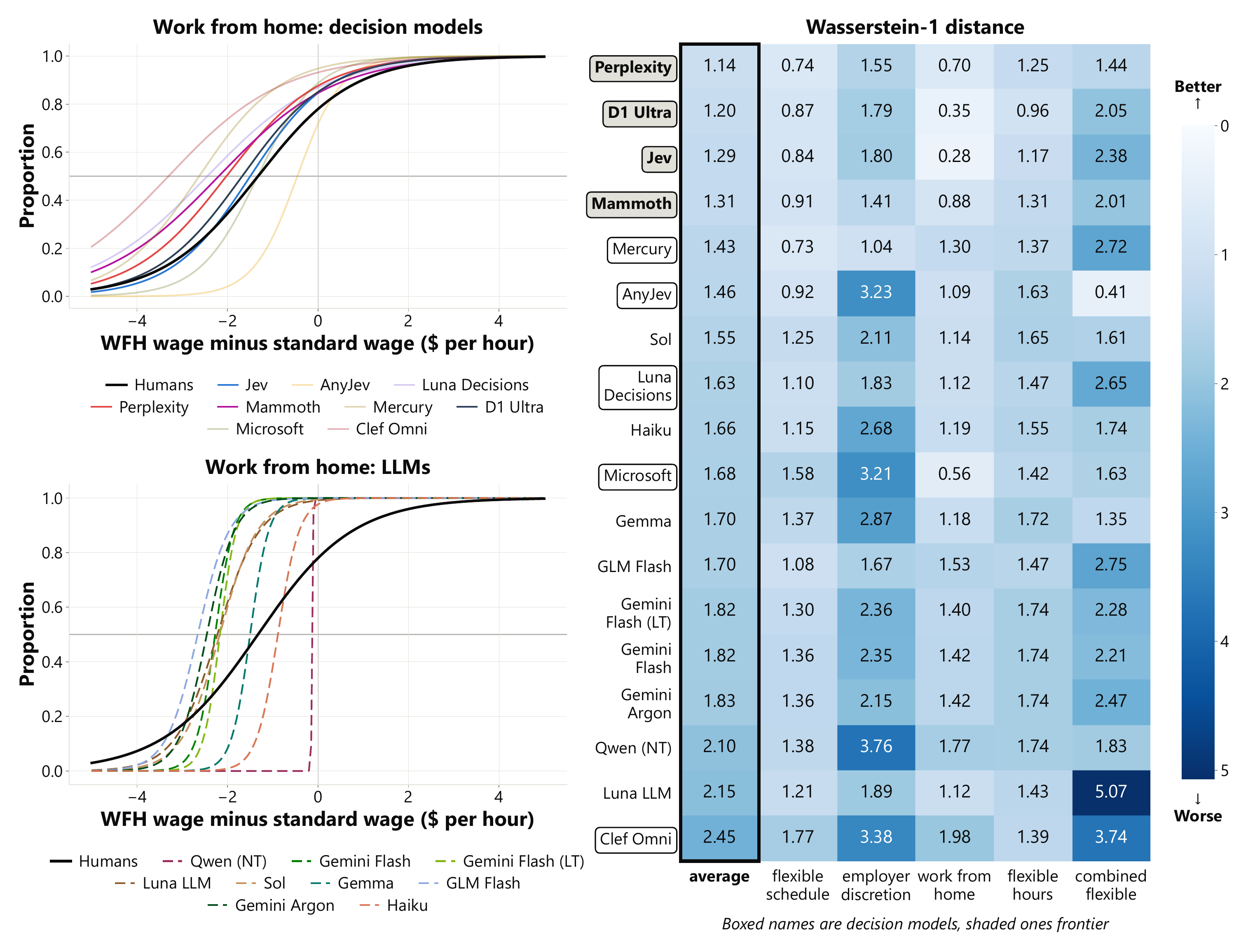}
    \caption{\textbf{Frontier DMs better capture the smooth curve in the human results compared to LLMs on labour choices.} Left panel shows DM results for the Work From Home treatment on top and the LLM results below, with the human data in black. The right panel shows the Wasserstein-1 distance between models and the human results across all 5 treatments. DMs are circled, with frontier DMs in bold. The full fitted curves for all five treatments are in Appendix \ref{app:labour}. Qwen (NT) is Qwen with no thinking, and Gemini Flash (LT) is Gemini Flash with low thinking.}
    \label{fig:mas}
\end{figure}

Figure \ref{fig:mas} shows that frontier DMs outperform LLMs and other DMs on average in representing human responses on the labour field experiment of \citet{mas2017valuing}. Bootstrapped two-sided tests at 1\% with Holm correction show that frontier DMs have statistically significantly lower mean Wasserstein-1 distances (averaged over the five treatments) than any LLM. The left panel from Figure \ref{fig:mas} shows why: DMs tend to capture the smooth increase in proportion of humans choosing one job over another as the wage gap increases. In contrast, LLMs look much more like step functions, as once they decide that one option is better than another, they tend to repeat their choice consistently. Details of the statistical tests and graphs of the other treatment curves are in Appendix \ref{app:labour}.

We omit Laya from this analysis as it does not give an upward sloping curve and has an inattention of near 100\%, in keeping with our earlier results that Laya is roughly as good as a random uniform distribution at predicting human responses. However, AnyJev does better in these experiments than in the consumption survey and economic games: it generally gives a smooth curve which often matches the human data well (mean Wasserstein-1 distance of 1.46 vs best DM at 1.14 and best LLM at 1.55). 

The above analysis compares the inattention-corrected curves between models and humans. \citet{mas2017valuing} estimated that around 25-30\% of the humans in their experiments were inattentive and chose at random. However, DMs and LLMs, except Laya, all have low estimated inattention proportions, all less than 10\% averaged over the five treatments with most near 0\% (see Appendix \ref{app:labour}).  It is unsurprising that DMs and LLMs do not reproduce human inattention as these models have been fine-tuned for use in cases, such as customer ticket prioritisation, where inattention would be highly undesirable. However, the low inattention of DMs and LLMs is worth bearing in mind if considering using these models as human surrogates.

The ability of DMs to react sensibly and smoothly to small differences in prompts, in this case small wage differences, is critical to being able to use DMs as human surrogates. Since DMs do not reason in the same way that LLMs do, they need an \enquote{intuitive} understanding of numbers and this experiment shows that they are indeed able to give sensible results with numerical data, even though they cannot do extended reasoning.

\subsection{Computational efficiency and its implications}
\label{sec:speed_cost}

\begin{figure}
    \centering
    \includegraphics[width=1\linewidth]{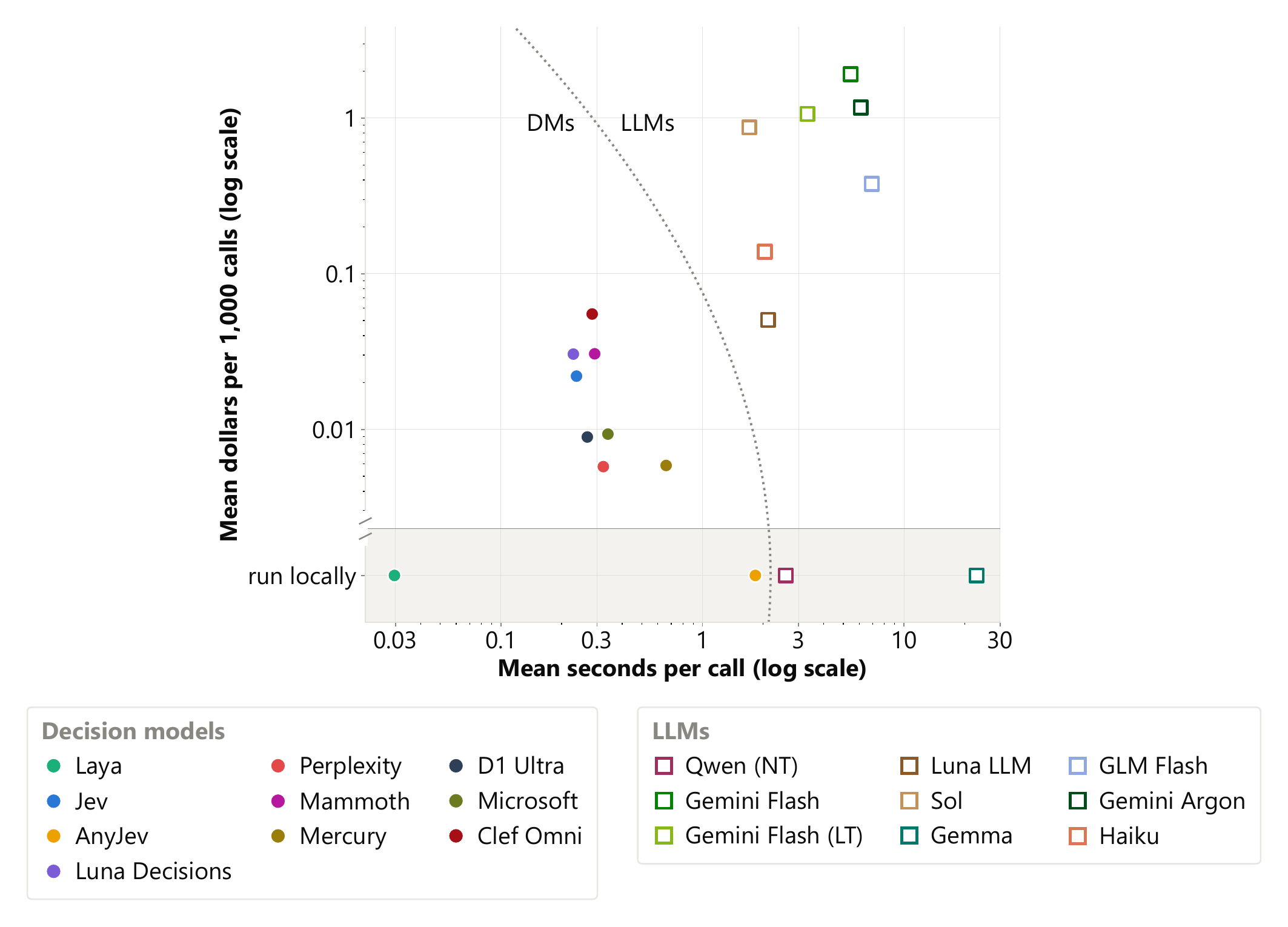}
    \caption{\textbf{DMs are faster and cheaper than LLMs.} Averaged across all calls to the models across the three experiments with equal weight to each experiment. Mean seconds per call and mean dollars per 1000 calls are on log scales. Qwen (with thinking) is omitted as it was only run for the economic games, where its average time per call was 657 seconds and cost was 0 as it was run locally. Gemma is included in the locally run models, but was in fact run using an API, where it is free to use. Qwen (NT) is Qwen with no thinking, while Gemini Flash (LT) is Gemini Flash with low thinking.}
    \label{fig:speed_vs_cost}
\end{figure}

DMs are faster and cheaper than LLMs; this holds across the three settings we consider and Figure \ref{fig:speed_vs_cost} illustrates this with the mean seconds per call and cost per 1000 calls across all experiments. Focusing only on the frontier DMs which give the most realistic representations of humans, a DM like Jev is around 10-100 times faster across experiments compared to the LLMs we test. Similarly, the cheapest frontier DM, Perplexity, is around 10-300 times cheaper than the LLMs we test (excluding locally-run LLMs, which are \enquote{free}). However, Jev is only around 2 times cheaper than Luna LLM and these numbers are subject to evolution as pricing has already changed for several DMs since their release. Nonetheless, our results illustrate that frontier DMs give a significant boost in terms of both speed and cost compared to LLMs. Results on speed and cost for all models and all experiments are reported in Appendix \ref{app:speed_cost}.

These speed and cost advantages transfer into increasing the feasible size and speed of in silico surveys and simulations. In consumer research, LLMs have been used to simulate purchase intent for product concepts \citep{maier2025llms}. Because LLMs asked directly for a rating produce unrealistic response distributions, these studies elicited free text and mapped it onto a rating scale, whereas a DM assigns a probability to each rating in a single call. Assuming that the average costs and speeds in our experiments translate to other economic settings, recreating the 18,600 decisions in \citet{maier2025llms} would take under 10 minutes with DMs while LLMs take around 1-3 hours.

Just as LLMs have already started to be used in economic ABMs \citep{li2024econagent, papadopoulos2026equations}, DMs offer an alternative for the agents in ABMs. Existing ABMs could be transformed to use DMs for decisions such as which occupations workers apply to \citep{del2021occupational}, how suppliers ration scarce output \citep{pichler2022forecasting}, or how firms set markups \citep{poledna2023economic}.  If we assume a simulation with 10,000 agents over 50 periods and a concurrency of 10 requests for all models, then in these ABMs, Perplexity would take around 4.5-13.5 hours and cost \$3-9, against Luna or Haiku taking over a day for \$25-70 and Sol or Argon at 1-4 days for \$400-600 (see Appendix \ref{app:speed_cost}). If techniques such as archetypes were also used, then a simulation such as that in \citet{chopra2024limits} could feasibly be scaled up from 8.4 million agents to around 80 million agents. Frontier DMs therefore offer a significant advantage compared to LLMs in scaling up economic surveys and computational simulations with AI agents.

\section{Discussion and Conclusion}
\label{sec:conclusion}

We examine whether Decision Models can act as human surrogates in economic research, comparing 10 DMs and 8 LLMs with human data from a household spending survey, classic economic games and a labour-market field experiment. Our experiments show that frontier DMs do a better job of representing the distribution of human behaviours than LLMs, and they do so in fractions of the time and cost. In the spending survey, frontier DMs represent the full human distribution better than LLMs without losing the ability to capture the modal human answer, while on labour choices they better capture the smooth curves seen in the human data. In the economic games, LLMs better capture the modal human answer than frontier DMs, but frontier DMs are still closer to humans on full distributional measures. Together, our results show that frontier DMs offer a scalable human surrogate which better captures the variation in human data.

While these results are promising, there are limitations to the experiments and data selected. The consumption and labour study data are US based. Moreover, since both LLMs and DMs were trained after the publication of the datasets and experiments used, there is likely data contamination. These limitations are partly addressed by the fact that the experiments in \citet{mei2024turing} include data from outside the US, and that LLM predictions of experimental treatment effects can remain accurate for studies published after the model’s training cut-off \citep{ashokkumar2026large}. Further, training after data publication does not guarantee that an LLM will replicate human behaviour \citep{saxena2026machine}. Nonetheless, validating these models against unpublished, out-of-distribution choice experiments and culturally diverse populations remains essential.

As Decision Models continue to evolve, they present opportunities to develop more faithful human surrogates. For example, models such as NumericJev already support fine-grained numerical outputs alongside discrete choices \citep{ye2026numericjev}. However, general advances in model capabilities do not always produce better human surrogates; frontier models often become hyper-rational on economic benchmarks, making them less representative of boundedly rational humans \citep{saxena2026machine}. Rather than relying on general model scaling, future work should fine-tune DMs explicitly on empirical human choice data, adapting behavioural calibration techniques previously applied to LLMs \citep{binz2025foundation, kolluri2025finetuning}. Because DMs have a modest parameter footprint and incur negligible inference costs, fine-tuning them specifically for behavioural fidelity offers a practical, scalable foundation for population-level economic simulation.

By addressing both the population-level scaling and restricted behavioural heterogeneity challenges of LLMs, Decision Models enable better large-scale computational simulations. While computational infeasibility has constrained LLM-based simulations, our results suggest that frontier DMs run approximately 10 to 100 times faster than LLMs. This efficiency allows existing agent-based simulations to scale by an order of magnitude, for example, expanding populations from 1 million to 10 million agents, or compressing the execution time of complex macroeconomic models from days to hours. Achieving 1-to-1 agent representations at a city or national scale at highly accessible costs enables practical applications across both public policy, such as modelling labour reallocation and supply-chain shocks, and commercial operations, where DMs can efficiently simulate heterogeneous consumer demand. In an era where AI is rapidly reshaping society, leveraging computer science to build computationally tractable and behaviourally faithful human surrogates is a promising tool for socio-economic modelling and policymaking.

\section{Acknowledgements}

We acknowledge Google for providing API credits for use in running Gemini experiments.

\section{Disclosure of AI Usage}

The authors declare the use of generative AI in the research and writing process. According to the GAIDeT taxonomy (2025), the following tasks were completed with assistance from GAI under full human supervision: code generation \& optimisation, proofreading \& editing. The tool used was Claude Opus 5.5 (Anthropic). After using these tools, the authors fully reviewed and edited all code and material.  Responsibility for the final manuscript lies entirely with the authors.

\bibliographystyle{plainnat}   
\bibliography{references}    

\newpage
\appendix

\section{Methods}

\subsection{Experiments}
\label{app:experiments}

\subsubsection*{Wording for the Household Spending Survey:}

\begin{itemize}
    \item Suppose next year you were to find your household with 10\% more income than you currently expect. What would you do with the extra income?
    \begin{enumerate}
        \item Save or invest all of it
        \item Spend or donate all of it
        \item Use all of it to pay down debts
        \item Spend some and save some
        \item Spend some and use part of it to pay down debts
        \item Save some and use part of it to pay down debts
        \item Spend some, save some and use some to pay down debts
    \end{enumerate}
    \item Imagine that next year you were to find yourself with 10\% less household income. What would you do?
    \begin{enumerate}
        \item Cut spending by the whole amount
        \item Not cut spending at all, but cut my savings by the whole amount
        \item Not cut spending at all, but increase my debt by borrowing the whole amount
        \item Cut spending by some and cut savings by some
        \item Cut spending by some and increase debt by some
        \item Cut savings by some and increase debt by some
        \item Cut spending by some, cut savings by some and increase debt some
    \end{enumerate}
\end{itemize}

\subsubsection*{Brief explanation of each game, taken verbatim from \citet{mei2024turing}:}

\begin{itemize}
    \item A Dictator Game - given an endowment of money, one player (the dictator) chooses how much of the money to keep and how much to donate to a second player. This involves altruism \citep{guth1982experimental, forsythe1994fairness}.
    \item An Ultimatum Game - given an endowment of money, one player (the proposer) offers a split of the money to a second player (the responder) who either accepts the split or rejects it in which case neither player gets anything. This involves fairness and spite \citep{guth1982experimental}.
    \item A Trust Game — given an endowment of money, one player (the investor) decides how much of the money to keep and passes the remainder to a second player (the banker), which is then tripled. The banker decides how much of that tripled revenue to keep and returns the remainder to the investor. This involves trust, fairness, altruism, and reciprocity \citep{berg1995trust}.
    \item A Bomb Risk Game — a player chooses how many boxes out of 100 to open and the player is rewarded for each opened box but loses everything if a randomly placed bomb is encountered. This involves risk aversion \citep{crosetto2013bomb}.
    \item A Public Goods Game — given an endowment of money, a player chooses how much of the money to keep and how much to contribute to a public good and receives half of the total amount donated to the public good by all four players. This involves free-riding, altruism, and cooperation \citep{andreoni1995cooperation}.
\end{itemize}

\subsubsection*{How to fit inattention and WTP logistic curves in \citet{mas2017valuing}:}

Let $g_i$ be position $A$'s hourly wage minus position $B$'s in observation $i$. Let $y_i$ be $1$ if $A$ was chosen and $0$ otherwise; for decision models, which return probabilities, $y_i$ is the probability of choosing $A$.

\paragraph{Inattention.}
Following \citet{mas2017valuing}, we allow a share of choices to ignore the
wage gap. We regress $y_i$ on $g_i$ by OLS using only gaps of \$2 or more,
$y_i = \beta_0 + \beta_1 g_i + \varepsilon_i$. Inattention is the
predicted share \emph{not} choosing $A$ when it pays \$5 more:
\[
  \hat\alpha = 1 - (\hat\beta_0 + 5\hat\beta_1),
\]
clipped to $[0, 0.49]$.

\paragraph{Curve.}
Given $\hat\alpha$, the probability of choosing $A$ is
\[
  P(A \mid g) = \hat\alpha + (1 - 2\hat\alpha)\,\Lambda(b g + c),
  \qquad \Lambda(x) = \frac{1}{1 + e^{-x}} .
\]
We estimate $b$ and $c$ by maximum likelihood, maximising $\sum_i \big[ y_i \log P(A \mid g_i) + (1 - y_i) \log\big(1 - P(A \mid g_i)\big) \big]$ by Fisher scoring with step halving. For plotting, we plot inattention-corrected curves using the above inattention-corrected fitted parameters.

\newpage
\subsection{Models}
\label{app:models}

\begin{table*}[htbp]
\centering
\label{tab:models}
\begin{tabular}{@{}l l l l l l p{4.6cm}@{}}
\toprule
Model Name & Model & Provider & Cost \$/M in/out & Device \\
\midrule
Jev & \texttt{jev-1.13.0} & TypeSafe & 0.042 / 0.00 & API \\
Perplexity & \texttt{pplx-decider-v1.1-27b} & Perplexity & 0.02 / 0.00 & API \\
Mammoth & \texttt{baseten-mammoth-27b} & Baseten & 0.10 / 0.00 & API \\
D1 Ultra & \texttt{d1-ultra} & Liquid & 0.042 / 0.00 & API \\
Luna Decisions & \texttt{gpt-6-luna} & OpenAI & 0.10 / 0.00 & API \\
Mercury & \texttt{mercury-decide} & Inception & 0.02 / 0.00 & API \\
Microsoft Decision 1 & \texttt{microsoft-decision-1} & Microsoft & 0.042 / 0.00 & API \\
Clef Omni & \texttt{clef-omni} & Cloudflare & 0.15 / 0.00 & API \\
AnyJev & \texttt{Qwen3.5-4B} & Nokia & - & GPU \\
Laya & \texttt{laya} & ConvAI & - & GPU \\
\bottomrule
\end{tabular}
\caption{\textbf{Decision Models used.} The GPU used was a Quadro RTX 6000.}
\end{table*}

\begin{table*}[htbp]
\centering
\label{tab:models}
\begin{tabular}{@{}l l l l l l p{4.6cm}@{}}
\toprule
Model Name & Model & Provider & Cost \$/M in/out & Device \\
\midrule
Gemini Flash & \texttt{gemini-3.8-flash} & Google & 0.75 / 3.75 & API\\
Gemini Argon & \texttt{gemini-4-argon} & Google & 2.0 / 10.0 & API\\
Gemma & \texttt{gemma-4-26b-a4b-it} & Google & - & API\\
Luna LLM & \texttt{gpt-6-luna} & OpenAI & 0.1 / 0.5 & API\\
Sol & \texttt{gpt-6-sol} & OpenAI & 2.0 / 10.0 & API\\
Haiku & \texttt{claude-haiku-5-5} & Anthropic & 0.1 / 0.5 & API\\
GLM Flash & \texttt{glm-5.3-flash} & Z.AI & 0.15 / 0.50 & API \\
Qwen & \texttt{Qwen/Qwen3.5-4B} & Alibaba & - & GPU \\
\bottomrule
\end{tabular}
\caption{\textbf{LLMs used.} APIs are all the provider's standard API interface. The GPU used was a Quadro RTX 6000 except for Qwen with thinking, where a mixture of a Quadro RTX 6000 and an Apple M4 were used to increase the speed of the runs. The Quadro RTX 6000 averaged 361 seconds per call and the M4 around 713 on similar length outputs. Both are several orders of magnitude slower than any of the DMs.}
\end{table*}

Gemini-3.8-Flash and GPT-6-Luna were run with their default settings, except when changing Gemini-3.8-Flash's reasoning to \enquote{low}. Qwen3.5-4B was run using Hugging Face \texttt{transformers}, bfloat16. With thinking on, the settings were: temperature 1.0, top-$p$ 0.95, top-$k$ 20 (as recommended). With thinking off, the settings were: temperature 0.7, top-$p$ 0.8, top-$k$ 20 (again according to the model card recommendation).

\subsection*{DM benchmarks}

We use the Artificial Analysis Decision Index to judge if a Decision Model is at the frontier of decision tasks. Not all models have been tested by Artificial Analysis; the scores, where available, are below:

\begin{table*}[htbp]
\centering
\label{tab:models}
\begin{tabular}{l c}
\toprule
Model Name & Artificial Analysis Decision Index \\
\midrule
Jev & 57 \\
Perplexity & 56 \\
Mammoth & 65.9 \\
D1 Ultra & 65  \\
\midrule
Luna Decisions & 48 \\
Mercury & 53 \\
Microsoft Decision 1 & 54 \\
Clef Omni & 44 \\
AnyJev & Not run \\
Laya & Not run \\
\bottomrule
\end{tabular}
\caption{\textbf{Decision Index scores.} Jev score is in fact Jev 1.14, a slightly newer version than we use. Artificial Analysis have not run Jev 1.13 on their benchmark, but TypeSafe, the creators of Jev, find little difference in score between the two models on their own benchmark. We therefore make an assumption that the same holds true on the Artificial Analysis benchmark. The results for Mammoth are those reported by the creators of Mammoth rather than those reported by Artificial Analysis (since they have not been run by Artificial Analysis).}
\end{table*}

\subsection*{LLM prompting:} 

We follow \citet{mei2024turing} and ask LLMs to reply with: \enquote{Please give only one choice, and highlight the position you choose with [] (such as [Position 1]).} For consistency with \citet{mei2024turing}, LLMs are given the system prompt: \enquote{You are a helpful assistant.}

\newpage

\section{Results}
\label{app:results}

\subsection{Consumer spending}
\label{app:consumer}

\begin{table}[htbp]
\centering
\begin{tabular}{lrrrr}
\toprule
LLM & Jev & Perplexity & Mammoth & D1 Ultra \\
\midrule
\multicolumn{5}{l}{\textit{Panel A: KL divergence from humans (bits)}} \\
\midrule
Qwen (NT) & $2.08^{*}$ & $2.04^{*}$ & $2.23^{*}$ & $2.13^{*}$ \\
Gemini Flash & $0.89^{*}$ & $0.84^{*}$ & $1.03^{*}$ & $0.93^{*}$ \\
Gemini Flash (LT) & $0.84^{*}$ & $0.79^{*}$ & $0.98^{*}$ & $0.88^{*}$ \\
Luna LLM & $1.23^{*}$ & $1.18^{*}$ & $1.37^{*}$ & $1.27^{*}$ \\
Sol & $1.54^{*}$ & $1.49^{*}$ & $1.68^{*}$ & $1.58^{*}$ \\
Gemma & $2.13^{*}$ & $2.08^{*}$ & $2.27^{*}$ & $2.17^{*}$ \\
GLM Flash & $0.17^{*}$ & $0.12^{*}$ & $0.31^{*}$ & $0.21^{*}$ \\
Gemini Argon & $0.17^{*}$ & $0.12^{*}$ & $0.31^{*}$ & $0.21^{*}$ \\
Haiku & $2.08^{*}$ & $2.03^{*}$ & $2.22^{*}$ & $2.12^{*}$ \\
\multicolumn{5}{l}{\footnotesize Decision model closer: 36; LLM closer: 0; no significant difference: 0.} \\
\midrule
\multicolumn{5}{l}{\textit{Panel B: Turing test win rate (\%)}} \\
\midrule
Qwen (NT) & $10.8^{*}$ & $13.9^{*}$ & $8.4$ & $12.8^{*}$ \\
Gemini Flash & $4.7$ & $7.7^{*}$ & $2.2$ & $6.6$ \\
Gemini Flash (LT) & $6.2$ & $9.2^{*}$ & $3.8$ & $8.1$ \\
Luna LLM & $0.3$ & $3.3$ & $-2.2^{*}$ & $2.2$ \\
Sol & $-8.8^{*}$ & $-5.8^{*}$ & $-11.3^{*}$ & $-6.9^{*}$ \\
Gemma & $5.0$ & $8.0$ & $2.5$ & $6.9$ \\
GLM Flash & $-3.1^{*}$ & $-0.0$ & $-5.5^{*}$ & $-1.1$ \\
Gemini Argon & $3.2$ & $6.3^{*}$ & $0.8$ & $5.2$ \\
Haiku & $2.7$ & $5.7$ & $0.3$ & $4.6$ \\
\multicolumn{5}{l}{\footnotesize Decision model closer: 6; LLM closer: 7; no significant difference: 23.} \\
\bottomrule
\end{tabular}
\caption{\textbf{Paired bootstrap tests of each DM against each LLM, averaged over the two SCE spending questions.} Panel A tests which model has a lower KL divergence while Panel B tests which model has a higher Turing Test score. Each cell is in bits (Panel A) or percentage points (Panel B), averaged over the 2 questions with equal weights: positive if the decision model is lower in KL divergence, higher in Turing Test. $^{*}$ Significant at the 1\% level (two-sided paired bootstrap test, 10,000 draws, Holm-corrected over each panel's tests). Each draw resamples the 1,000 personas with replacement.}
\end{table}

\begin{table}[htbp]
\centering

\label{tab:sce_bootstrap_uniform}
\begin{tabular}{lrr}
\toprule
Decision model & KL divergence from humans (bits) & Turing test (\%) \\
\midrule
Jev & $0.28^{*}$ & $12.4^{*}$ \\
D1 Ultra & $0.32^{*}$ & $14.3^{*}$ \\
Perplexity & $0.23^{*}$ & $15.4^{*}$ \\
Mammoth & $0.42^{*}$ & $9.9^{*}$ \\
\bottomrule
\end{tabular}
\caption{\textbf{Paired bootstrap tests of the frontier DMs against the uniform distribution, averaged over the two SCE spending questions.} Values are as above, with a higher number meaning DMs have a lower KL divergence or higher Turing test scores. $^{*}$ Significant at the 1\% level (two-sided paired bootstrap test, 10,000 draws, Holm-corrected). Each draw resamples the 1,000 personas with replacement.}
\end{table}

\subsection*{Robustness checks}

The first robustness check looks at changing the wording for the question asking for a response after a 10\% reduction in earnings. This is considered because in the original survey the two questions are asked one after the other. However, since the DMs and LLMs get each question separately, the exact survey wording for the second question no longer made logical sense, and had to be slightly changed for these experiments. We therefore consider two versions of the prompt, the first of which is the prompt we use throughout the paper and the second is the robustness check:

\begin{enumerate}
    \item \enquote{Imagine that next year you were to find yourself with 10\% less household income. What would you do?}
    \item \enquote{Suppose next year you were to find your household with 10\% less income than you currently expect. What would you do?}
\end{enumerate}

The second check considers adding the instruction to \enquote{Act like a human} before prompts for both DMs and LLMs. The third robustness check considers only LLMs. LLMs are prompted in the same way as in \citet{mei2024turing}. However, this includes adding a system prompt of \enquote{You are a helpful assistant}. We look at taking this system prompt out for LLMs.

\begin{figure}
    \centering
    \includegraphics[width=1\linewidth]{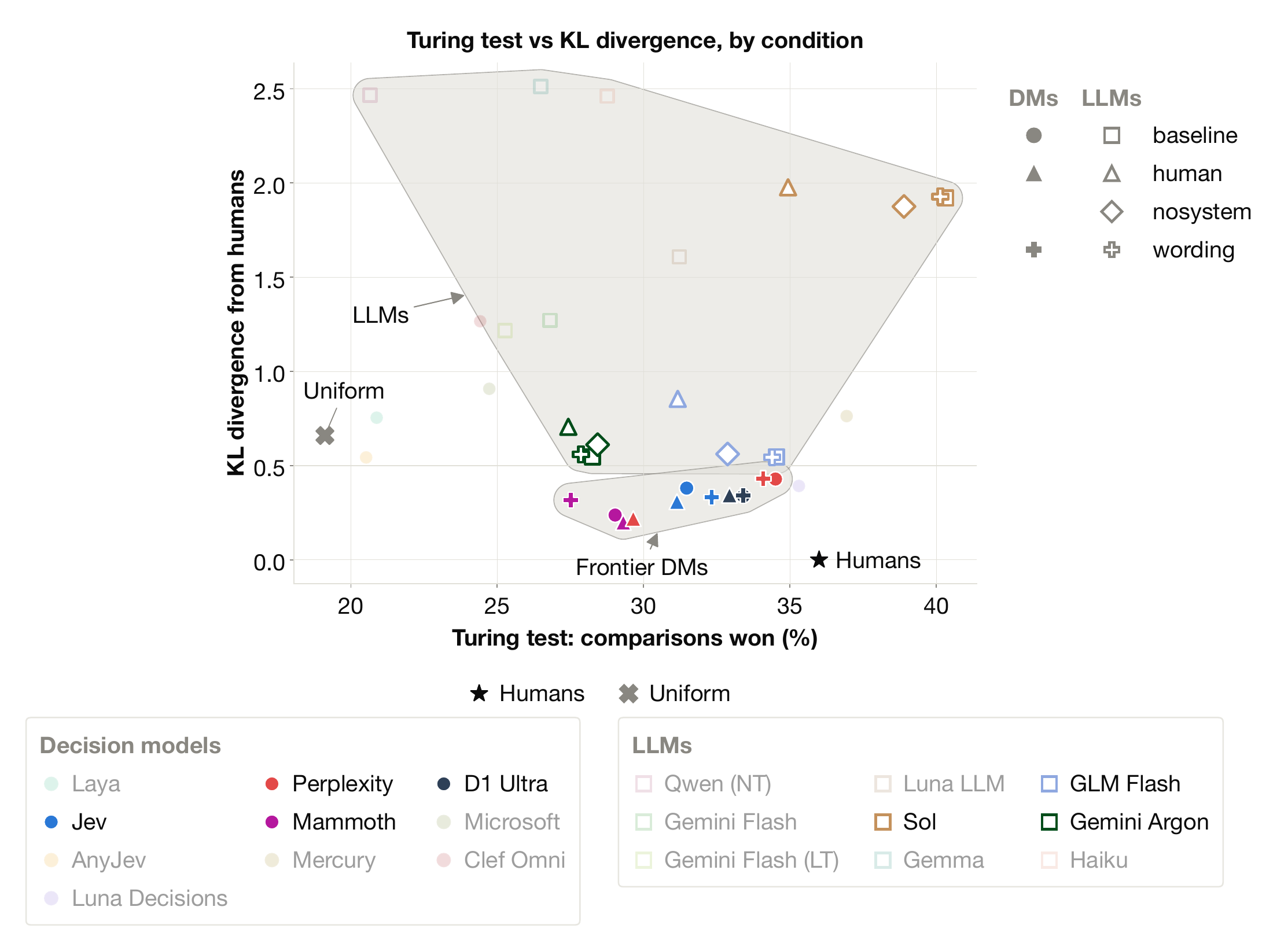}
    \caption{\textbf{Our results are robust to three different robustness checks.} DMs are solid while LLMs are hollow.  Human is the test where models are told to act like a human. Nosystem removes the system prompt for the LLMs while wording changes the wording for the loss of income question. Scores are averaged over the two consumer spending survey questions with equal weight.}
    \label{fig:app:robustness}
\end{figure}

We find that our results are very robust to these three considered changes, with frontier DMs still closer to humans on KL divergence and mixed results with the Turing Test. Figure \ref{fig:app:robustness} shows the results.

\newpage

\subsection{Economic Games}
\label{app:economicgames}

\begin{table}[htbp]
\centering
\begin{tabular}{lrrrr}
\toprule
LLM & Jev & Perplexity & Mammoth & D1 Ultra \\
\midrule
\multicolumn{5}{l}{\textit{Panel A: Wasserstein-1 distance (\% of range)}} \\
\midrule
Qwen & $1.0$ & $10.5^{*}$ & $9.5^{*}$ & $9.4^{*}$ \\
Qwen (NT) & $9.8^{*}$ & $19.4^{*}$ & $18.3^{*}$ & $18.3^{*}$ \\
Gemini Flash & $8.3^{*}$ & $17.9^{*}$ & $16.9^{*}$ & $16.8^{*}$ \\
Gemini Flash (LT) & $10.3^{*}$ & $19.9^{*}$ & $18.9^{*}$ & $18.8^{*}$ \\
Luna LLM & $2.7^{*}$ & $12.2^{*}$ & $11.2^{*}$ & $11.1^{*}$ \\
Sol & $9.3^{*}$ & $18.9^{*}$ & $17.8^{*}$ & $17.8^{*}$ \\
Gemma & $6.1^{*}$ & $15.7^{*}$ & $14.6^{*}$ & $14.5^{*}$ \\
GLM Flash & $5.4^{*}$ & $15.0^{*}$ & $14.0^{*}$ & $13.9^{*}$ \\
Gemini Argon & $7.6^{*}$ & $17.1^{*}$ & $16.1^{*}$ & $16.0^{*}$ \\
Haiku & $3.1^{*}$ & $12.6^{*}$ & $11.6^{*}$ & $11.5^{*}$ \\
\multicolumn{5}{l}{\footnotesize Decision model closer: 39; LLM closer: 0; no significant difference: 1.} \\
\midrule
\multicolumn{5}{l}{\textit{Panel B: Turing test win rate (\%)}} \\
\midrule
Qwen & $-28.0^{*}$ & $-29.1^{*}$ & $-28.0^{*}$ & $-27.1^{*}$ \\
Qwen (NT) & $-9.0^{*}$ & $-10.0^{*}$ & $-9.0^{*}$ & $-8.1^{*}$ \\
Gemini Flash & $-20.9^{*}$ & $-22.0^{*}$ & $-20.9^{*}$ & $-20.0^{*}$ \\
Gemini Flash (LT) & $-22.0^{*}$ & $-23.0^{*}$ & $-21.9^{*}$ & $-21.1^{*}$ \\
Luna LLM & $-22.6^{*}$ & $-23.6^{*}$ & $-22.5^{*}$ & $-21.6^{*}$ \\
Sol & $-21.3^{*}$ & $-22.3^{*}$ & $-21.2^{*}$ & $-20.3^{*}$ \\
Gemma & $-24.9^{*}$ & $-25.9^{*}$ & $-24.8^{*}$ & $-23.9^{*}$ \\
GLM Flash & $-21.4^{*}$ & $-22.4^{*}$ & $-21.3^{*}$ & $-20.4^{*}$ \\
Gemini Argon & $-17.9^{*}$ & $-19.0^{*}$ & $-17.9^{*}$ & $-17.0^{*}$ \\
Haiku & $-21.6^{*}$ & $-22.7^{*}$ & $-21.6^{*}$ & $-20.7^{*}$ \\
\multicolumn{5}{l}{\footnotesize Decision model closer: 0; LLM closer: 40; no significant difference: 0.} \\
\bottomrule
\end{tabular}
\caption{\textbf{Bootstrap tests of frontier DMs, using the choice answer, against each LLM, averaged over \citet{mei2024turing} games.} Each cell is how much lower the Wasserstein-1 distance or higher the Turing Test the decision model is than the LLM, in percentage points, averaged over the 7 games with equal weights: positive if the decision model is lower Wasserstein-1 or higher Turing test. $^{*}$ Significant at the 1\% level (two-sided bootstrap test, 10,000 draws, Holm-corrected over each panel's tests). Each model's calls are resampled with replacement in each game for the bootstrap.}
\end{table}

\begin{table}[htbp]
\centering
\begin{tabular}{lrrrr}
\toprule
LLM & Jev & Perplexity & Mammoth & D1 Ultra \\
\midrule
\multicolumn{5}{l}{\textit{Panel A: Wasserstein-1 distance (\% of range)}} \\
\midrule
Qwen & $5.6^{*}$ & $9.1^{*}$ & $9.2^{*}$ & $7.5^{*}$ \\
Qwen (NT) & $14.4^{*}$ & $17.9^{*}$ & $18.1^{*}$ & $16.3^{*}$ \\
Gemini Flash & $13.0^{*}$ & $16.4^{*}$ & $16.6^{*}$ & $14.9^{*}$ \\
Gemini Flash (LT) & $15.0^{*}$ & $18.4^{*}$ & $18.6^{*}$ & $16.8^{*}$ \\
Luna LLM & $7.3^{*}$ & $10.8^{*}$ & $10.9^{*}$ & $9.2^{*}$ \\
Sol & $13.9^{*}$ & $17.4^{*}$ & $17.6^{*}$ & $15.8^{*}$ \\
Gemma & $10.7^{*}$ & $14.2^{*}$ & $14.4^{*}$ & $12.6^{*}$ \\
GLM Flash & $10.1^{*}$ & $13.5^{*}$ & $13.7^{*}$ & $12.0^{*}$ \\
Gemini Argon & $12.2^{*}$ & $15.7^{*}$ & $15.9^{*}$ & $14.1^{*}$ \\
Haiku & $7.7^{*}$ & $11.2^{*}$ & $11.4^{*}$ & $9.6^{*}$ \\
\multicolumn{5}{l}{\footnotesize Decision model closer: 40; LLM closer: 0; no significant difference: 0.} \\
\midrule
\multicolumn{5}{l}{\textit{Panel B: Turing test win rate (\%)}} \\
\midrule
Qwen & $-25.3^{*}$ & $-25.9^{*}$ & $-24.3^{*}$ & $-25.5^{*}$ \\
Qwen (NT) & $-6.3^{*}$ & $-6.9^{*}$ & $-5.3^{*}$ & $-6.5^{*}$ \\
Gemini Flash & $-18.2^{*}$ & $-18.8^{*}$ & $-17.2^{*}$ & $-18.4^{*}$ \\
Gemini Flash (LT) & $-19.3^{*}$ & $-19.8^{*}$ & $-18.3^{*}$ & $-19.5^{*}$ \\
Luna LLM & $-19.9^{*}$ & $-20.4^{*}$ & $-18.9^{*}$ & $-20.0^{*}$ \\
Sol & $-18.6^{*}$ & $-19.1^{*}$ & $-17.6^{*}$ & $-18.7^{*}$ \\
Gemma & $-22.2^{*}$ & $-22.7^{*}$ & $-21.2^{*}$ & $-22.3^{*}$ \\
GLM Flash & $-18.7^{*}$ & $-19.2^{*}$ & $-17.7^{*}$ & $-18.8^{*}$ \\
Gemini Argon & $-15.2^{*}$ & $-15.8^{*}$ & $-14.2^{*}$ & $-15.4^{*}$ \\
Haiku & $-18.9^{*}$ & $-19.5^{*}$ & $-17.9^{*}$ & $-19.1^{*}$ \\
\multicolumn{5}{l}{\footnotesize Decision model closer: 0; LLM closer: 40; no significant difference: 0.} \\
\bottomrule
\end{tabular}
\caption{\textbf{Bootstrap tests of frontier DMs, using the score answer, against each LLM, averaged over \citet{mei2024turing} games.} Each cell is how much lower the Wasserstein-1 distance or higher the Turing Test the decision model is than the LLM, in percentage points, averaged over the 7 games with equal weights: positive if the decision model is lower Wasserstein-1 or higher Turing test. $^{*}$ Significant at the 1\% level (two-sided bootstrap test, 10,000 draws, Holm-corrected over each panel's tests). Each model's calls are resampled with replacement in each game for the bootstrap.}
\end{table}

\begin{table}[htbp]
\centering
\begin{tabular}{lrrrr}
\toprule
LLM & Jev & Perplexity & Mammoth & D1 Ultra \\
\midrule
\multicolumn{5}{l}{\textit{Panel A: KL divergence (bits)}} \\
\midrule
Qwen & $2.24^{*}$ & $2.83^{*}$ & $2.75^{*}$ & $2.89^{*}$ \\
Qwen (NT) & $3.81^{*}$ & $4.40^{*}$ & $4.32^{*}$ & $4.46^{*}$ \\
Gemini Flash & $3.51^{*}$ & $4.10^{*}$ & $4.02^{*}$ & $4.16^{*}$ \\
Gemini Flash (LT) & $3.80^{*}$ & $4.38^{*}$ & $4.30^{*}$ & $4.44^{*}$ \\
Luna LLM & $2.66^{*}$ & $3.25^{*}$ & $3.16^{*}$ & $3.31^{*}$ \\
Sol & $3.55^{*}$ & $4.14^{*}$ & $4.05^{*}$ & $4.20^{*}$ \\
Gemma & $2.75^{*}$ & $3.34^{*}$ & $3.26^{*}$ & $3.40^{*}$ \\
GLM Flash & $2.88^{*}$ & $3.47^{*}$ & $3.38^{*}$ & $3.53^{*}$ \\
Gemini Argon & $3.81^{*}$ & $4.40^{*}$ & $4.32^{*}$ & $4.46^{*}$ \\
Haiku & $2.71^{*}$ & $3.29^{*}$ & $3.21^{*}$ & $3.35^{*}$ \\
\multicolumn{5}{l}{\footnotesize Decision model closer: 40; LLM closer: 0; no significant difference: 0.} \\
\bottomrule
\end{tabular}
\caption{\textbf{Bootstrap tests of frontier DMs, using the choice answer, against each LLM, averaged over \citet{mei2024turing} games.} Each cell is how much lower the KL divergence of the decision model is than the LLM, averaged over the 7 games with equal weights: positive if the decision model is lower KL divergence. $^{*}$ Significant at the 1\% level (two-sided bootstrap test, 10,000 draws, Holm-corrected). Each model's calls are resampled with replacement in each game for the bootstrap.}
\end{table}

\begin{table}[htbp]
\centering
\begin{tabular}{lrrrr}
\toprule
LLM & Jev & Perplexity & Mammoth & D1 Ultra \\
\midrule
\multicolumn{5}{l}{\textit{Panel A: KL divergence (bits)}} \\
\midrule
Qwen & $2.61^{*}$ & $2.95^{*}$ & $2.96^{*}$ & $2.93^{*}$ \\
Qwen (NT) & $4.18^{*}$ & $4.52^{*}$ & $4.53^{*}$ & $4.50^{*}$ \\
Gemini Flash & $3.88^{*}$ & $4.22^{*}$ & $4.23^{*}$ & $4.20^{*}$ \\
Gemini Flash (LT) & $4.16^{*}$ & $4.50^{*}$ & $4.52^{*}$ & $4.48^{*}$ \\
Luna LLM & $3.03^{*}$ & $3.37^{*}$ & $3.38^{*}$ & $3.35^{*}$ \\
Sol & $3.92^{*}$ & $4.26^{*}$ & $4.27^{*}$ & $4.24^{*}$ \\
Gemma & $3.12^{*}$ & $3.46^{*}$ & $3.47^{*}$ & $3.44^{*}$ \\
GLM Flash & $3.25^{*}$ & $3.59^{*}$ & $3.60^{*}$ & $3.57^{*}$ \\
Gemini Argon & $4.18^{*}$ & $4.52^{*}$ & $4.53^{*}$ & $4.50^{*}$ \\
Haiku & $3.07^{*}$ & $3.41^{*}$ & $3.42^{*}$ & $3.39^{*}$ \\
\multicolumn{5}{l}{\footnotesize Decision model closer: 40; LLM closer: 0; no significant difference: 0.} \\
\bottomrule
\end{tabular}
\caption{\textbf{Bootstrap tests of frontier DMs, using the score answer, against each LLM, averaged over \citet{mei2024turing} games.} Each cell is how much lower the KL divergence of the decision model is than the LLM, averaged over the 7 games with equal weights: positive if the decision model is lower KL divergence. $^{*}$ Significant at the 1\% level (two-sided bootstrap test, 10,000 draws, Holm-corrected). Each model's calls are resampled with replacement in each game for the bootstrap.}
\end{table}

\begin{table}[htbp]
\centering
\begin{tabular}{lrr}
\toprule
Decision model & Wasserstein-1 distance (\% of range) & Turing test (\%) \\
\midrule
Jev (choice) & $-7.2^{*}$ & $5.3^{*}$ \\
Jev (score) & $-2.5^{*}$ & $8.0^{*}$ \\
D1 Ultra (choice) & $1.3^{*}$ & $6.3^{*}$ \\
D1 Ultra (score) & $-0.7^{*}$ & $7.9^{*}$ \\
Perplexity (choice) & $2.4^{*}$ & $4.3^{*}$ \\
Perplexity (score) & $0.9^{*}$ & $7.5^{*}$ \\
Mammoth (choice) & $1.3^{*}$ & $5.4^{*}$ \\
Mammoth (score) & $1.1^{*}$ & $9.0^{*}$ \\
\bottomrule
\end{tabular}
\caption{\textbf{Bootstrap tests of frontier DMs against the uniform distribution.} Each cell is how much lower the Wasserstein-1 distance or higher the Turing Test the decision model is than the uniform distribution, averaged over the 7 games with equal weights: positive if the decision model is lower Wasserstein-1 or higher Turing test. $^{*}$ Significant at the 1\% level (two-sided bootstrap test, 10,000 draws, Holm-corrected). Each model's calls are resampled with replacement in each game for the bootstrap.}
\end{table}

\clearpage

\subsection{Labour choices}
\label{app:labour}

\begin{figure}[htbp]
    \centering
    \includegraphics[width=0.8\linewidth]{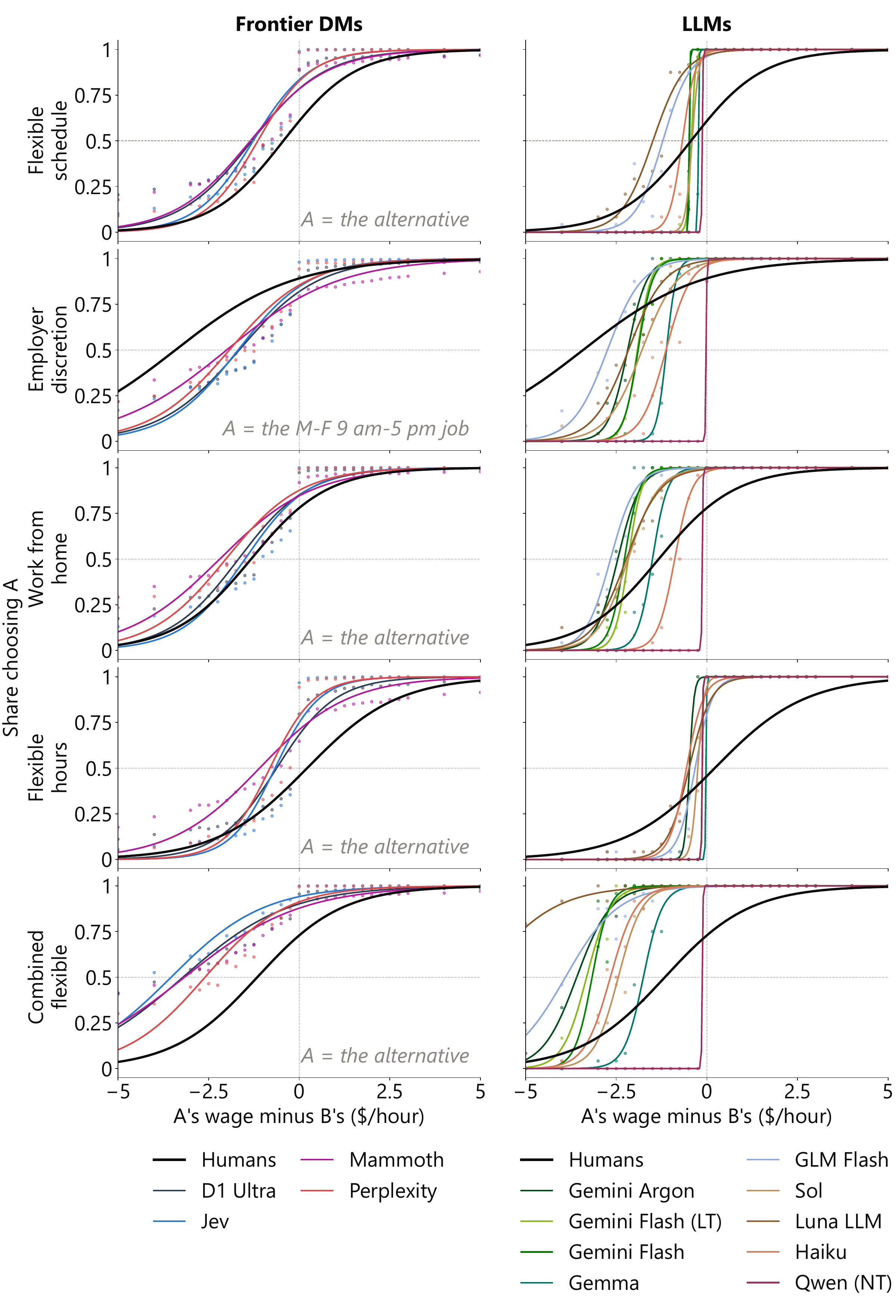}
    \caption{\textbf{Fitted inattention-corrected curves for all 5 treatments.} Job A is the more flexible alternative, so in 4 treatments it is the alternative and when the treatment is employer discretion, it is the standard 9-5 40 hour job. The left column shows frontier DMs, while the right column shows all LLMs.}
\end{figure}

\begin{table}[htbp]
\centering
\label{tab:mas_pallais_bootstrap}
\begin{tabular}{lrrrr}
\toprule
LLM & Jev & Perplexity & Mammoth & D1 Ultra \\
\midrule
Qwen (NT) & $0.80^{*}$ & $0.96^{*}$ & $0.79^{*}$ & $0.89^{*}$ \\
Gemini Flash & $0.52^{*}$ & $0.68^{*}$ & $0.51^{*}$ & $0.61^{*}$ \\
Gemini Flash (LT) & $0.52^{*}$ & $0.68^{*}$ & $0.51^{*}$ & $0.61^{*}$ \\
Luna LLM & $0.85^{*}$ & $1.01^{*}$ & $0.84^{*}$ & $0.94^{*}$ \\
Sol & $0.26^{*}$ & $0.42^{*}$ & $0.25^{*}$ & $0.35^{*}$ \\
Gemma & $0.40^{*}$ & $0.56^{*}$ & $0.39^{*}$ & $0.49^{*}$ \\
GLM Flash & $0.40^{*}$ & $0.56^{*}$ & $0.39^{*}$ & $0.49^{*}$ \\
Gemini Argon & $0.53^{*}$ & $0.69^{*}$ & $0.52^{*}$ & $0.62^{*}$ \\
Haiku & $0.37^{*}$ & $0.52^{*}$ & $0.35^{*}$ & $0.46^{*}$ \\
\midrule
\multicolumn{5}{l}{\footnotesize Decision model closer: 36; LLM closer: 0; no significant difference: 0.} \\
\bottomrule
\end{tabular}
\caption{\textbf{Bootstrap tests of frontier DMs against each LLM.} Each cell is how much closer to the applicants' fitted WTP distribution the decision model's is than the LLM's, in \$ per hour of Wasserstein-1 distance (the area between the fitted choice curves without inattention), averaged over the 5 treatments with equal weights: positive if the decision model is closer. $^{*}$ Significant at the 1\% level (two-sided bootstrap test, 10,000 draws, Holm-corrected over the tests).}
\end{table}

\begin{table}[htbp]
\centering
\label{tab:mas_pallais_inattention}
\begin{tabular}{lrrrrrr}
\toprule
 & FS & ED & WFH & FH & CF & Mean \\
\midrule
Laya & $0.49^{\dagger}$ & $0.49^{\dagger}$ & $0.43^{\dagger}$ & $0.49^{\dagger}$ & $0.41^{\dagger}$ & $0.46$ \\
Jev & $0.00$ & $0.00$ & $0.00$ & $0.00$ & $0.00$ & $0.00$ \\
AnyJev & $0.00$ & $0.03$ & $0.00$ & $0.01$ & $0.00$ & $0.01$ \\
Luna Decisions & $0.00$ & $0.01$ & $0.00$ & $0.00$ & $0.00$ & $0.00$ \\
Perplexity & $0.00$ & $0.01$ & $0.00$ & $0.00$ & $0.00$ & $0.00$ \\
Mammoth & $0.03$ & $0.07$ & $0.02$ & $0.08$ & $0.03$ & $0.05$ \\
Mercury & $0.00$ & $0.00$ & $0.00$ & $0.00$ & $0.00$ & $0.00$ \\
D1 Ultra & $0.03$ & $0.02$ & $0.01$ & $0.02$ & $0.01$ & $0.02$ \\
Microsoft & $0.00$ & $0.00$ & $0.00$ & $0.00$ & $0.00$ & $0.00$ \\
Clef Omni & $0.02$ & $0.02$ & $0.03$ & $0.02$ & $0.03$ & $0.02$ \\
\midrule
Qwen (NT) & $0.00$ & $0.00$ & $0.00$ & $0.00$ & $0.00$ & $0.00$ \\
Gemini Flash & $0.00$ & $0.00$ & $0.00$ & $0.00$ & $0.00$ & $0.00$ \\
Gemini Flash (LT) & $0.00$ & $0.00$ & $0.00$ & $0.00$ & $0.00$ & $0.00$ \\
Luna LLM & $0.00$ & $0.00$ & $0.00$ & $0.00$ & $0.00$ & $0.00$ \\
Sol & $0.00$ & $0.00$ & $0.00$ & $0.00$ & $0.00$ & $0.00$ \\
Gemma & $0.00$ & $0.00$ & $0.00$ & $0.00$ & $0.00$ & $0.00$ \\
GLM Flash & $0.00$ & $0.00$ & $0.00$ & $0.00$ & $0.00$ & $0.00$ \\
Gemini Argon & $0.00$ & $0.00$ & $0.00$ & $0.00$ & $0.00$ & $0.00$ \\
Haiku & $0.00$ & $0.00$ & $0.00$ & $0.00$ & $0.00$ & $0.00$ \\
\bottomrule
\end{tabular}
\caption{\textbf{Estimated inattention in each of \citet{mas2017valuing}'s treatments.} Inattention $\alpha$ in the inattention-corrected logit $P(A \mid \text{gap}) = \alpha + (1 - 2\alpha) F(b\,\text{gap} + c)$: the share not choosing A when it pays \$5 more, from a linear fit of the choices at wage gaps of \$2 to \$5. Comparison to the human average of 25-30\% then requires the above estimate of $\alpha$ to be doubled. Mean is the average over the 5 treatments with equal weights. $^{\dagger}$ Above 0.4, so no WTP is fitted. We use the following abbreviations to save space: FS (Flexible Schedule), ED (Employer Discretion), WFH (Work From Home), FH (Flexible Hours), CF (Combined Flexible).}
\end{table}

\clearpage
\subsection{Speed and cost analysis}
\label{app:speed_cost}

We estimate the cost and time of using DMs and LLMs as human surrogates in the applications discussed in Section \ref{sec:speed_cost}, using the cost and time per call of each model in Table~\ref{tab:speed_cost}. A simulation with $N$ calls to a model with cost $c$ and time $\tau$ per call costs $Nc$ and, with $K$ concurrent requests, takes $N\tau/K$ to run. We set $K = 10$ for all models. In an ABM, $N$ is the product of the number of agents, periods and calls per decision. Periods run in sequence, but each period contains $10^4$ decisions against $10$ concurrent requests, so the sequencing does not change the estimate. We assume that every agent decides in every period, which is an upper bound for models in which only some agents act, and that prompts are of similar length to those in our experiments. Calibration and Monte Carlo analysis multiply all figures by the number of runs.

\citet{maier2025llms} simulated purchase intent on a five-point rating scale. A DM returns a probability for each rating in one call. We consider a firm testing 57 products over a total of 9,300 human choices, giving 18,600 ratings since \citet{maier2025llms} repeat calls each time. For LLMs we assume one call per rating, which understates the cost of the method of \citet{maier2025llms}, as it generates free text and computes an embedding for each response.

For ABMs, we consider $10^4$ agents over 50 periods, giving $5 \times 10^5$ decisions, and assume that a DM chooses among at most ten options per call. In the occupational mobility model of \citet{del2021occupational}, workers choose which of 464 occupations to apply to. A DM can make this choice in three nested calls by arranging the occupations in a tree with at most ten branches per node ($8^3 = 512 \geq 464$). For LLMs we assume one call per decision. This favours LLMs: they make a third as many calls as DMs. In the input--output model of \citet{pichler2022forecasting}, suppliers that cannot meet demand ration their customers by a fixed rule, and \citet{pichler2022forecasting, pichler2022simultaneous} show that the estimated output losses depend on the rule used. Letting each supplier choose among a set of rules takes one DM call per period. Because \citet{pichler2022forecasting} model industries rather than firms, our figures correspond to a firm-level version of the model. In the model of \citet{poledna2023economic}, firms set prices each period, and choosing among at most ten discrete price changes also takes one DM call.

Table~\ref{tab:project_costs} reports the estimates of time and cost for frontier DMs and LLMs across the survey and ABMs considered. 

\begin{table}[htbp]
\centering
\setlength{\tabcolsep}{3pt}
\begin{tabular}{@{}lrrrrrrrr@{}}
\toprule
 & \multicolumn{4}{c}{Mean seconds per call} & \multicolumn{4}{c}{Mean \$ per 1,000 calls} \\
\cmidrule(lr){2-5} \cmidrule(lr){6-9}
 & Spending & Games & Labour & \textbf{Mean} & Spending & Games & Labour & \textbf{Mean} \\
\midrule
Laya & $0.03$ & $0.03$ & $0.03$ & $0.03$ & $0.000$ & $0.000$ & $0.000$ & $0.000$ \\
Jev & $0.26$ & $0.24$ & $0.21$ & $0.24$ & $0.024$ & $0.022$ & $0.020$ & $0.022$ \\
AnyJev & $2.81$ & $1.94$ & $0.75$ & $1.83$ & $0.000$ & $0.000$ & $0.000$ & $0.000$ \\
Luna Decisions & $0.22$ & $0.22$ & $0.24$ & $0.23$ & $0.033$ & $0.031$ & $0.028$ & $0.030$ \\
Perplexity & $0.33$ & $0.31$ & $0.33$ & $0.32$ & $0.006$ & $0.006$ & $0.005$ & $0.006$ \\
Mammoth & $0.24$ & $0.43$ & $0.21$ & $0.29$ & $0.033$ & $0.033$ & $0.026$ & $0.031$ \\
Mercury & $0.86$ & $0.71$ & $0.41$ & $0.66$ & $0.006$ & $0.007$ & $0.005$ & $0.006$ \\
D1 Ultra & $0.28$ & $0.27$ & $0.25$ & $0.27$ & $0.010$ & $0.008$ & $0.008$ & $0.009$ \\
Microsoft & $0.34$ & $0.35$ & $0.33$ & $0.34$ & $0.010$ & $0.009$ & $0.009$ & $0.009$ \\
Clef Omni & $0.33$ & $0.23$ & $0.29$ & $0.28$ & $0.059$ & $0.062$ & $0.044$ & $0.055$ \\
\midrule
Qwen & -- & $656.67$ & -- & -- & -- & $0.000$ & -- & -- \\
Qwen (NT) & $0.84$ & $6.21$ & $0.73$ & $2.59$ & $0.000$ & $0.000$ & $0.000$ & $0.000$ \\
Gemini Flash & $5.13$ & $7.23$ & $3.96$ & $5.44$ & $1.844$ & $2.634$ & $1.263$ & $1.914$ \\
Gemini Flash (LT) & $3.39$ & $3.65$ & $2.92$ & $3.32$ & $1.076$ & $1.333$ & $0.778$ & $1.062$ \\
Luna LLM & $1.67$ & $2.90$ & $1.79$ & $2.12$ & $0.050$ & $0.062$ & $0.039$ & $0.050$ \\
Sol & $1.46$ & $2.33$ & $1.34$ & $1.71$ & $0.910$ & $1.039$ & $0.664$ & $0.871$ \\
Gemma & $22.04$ & $33.95$ & $12.80$ & $22.93$ & $0.000$ & $0.000$ & $0.000$ & $0.000$ \\
GLM Flash & $3.58$ & $4.73$ & $12.49$ & $6.93$ & $0.327$ & $0.512$ & $0.296$ & $0.378$ \\
Gemini Argon & $3.83$ & $8.91$ & $5.57$ & $6.10$ & $0.801$ & $1.533$ & $1.174$ & $1.169$ \\
Haiku & $1.79$ & $2.28$ & $2.05$ & $2.04$ & $0.125$ & $0.151$ & $0.139$ & $0.139$ \\
\bottomrule
\end{tabular}
\caption{\textbf{Mean time and cost per call in each experiment.} Seconds exclude model loading and a warm-up call. Cost is worked out from each call's tokens and the API's list prices; models run locally cost 0, and Gemma is free. Mean is the average over the 3 experiments with equal weights, blank for a model not run in all of them.}
\label{tab:speed_cost}
\end{table}

\begin{table}[htbp]
\centering
\begin{tabular}{lrrrrrrrr}
\toprule
 & \multicolumn{2}{c}{Average} & \multicolumn{2}{c}{Purchases} & \multicolumn{2}{c}{Occ. mobility} & \multicolumn{2}{c}{Rationing \& Markup} \\
\cmidrule(lr){2-3} \cmidrule(lr){4-5} \cmidrule(lr){6-7} \cmidrule(lr){8-9}
 & Cost & Time & Cost & Time & Cost & Time & Cost & Time \\
\midrule
\multicolumn{9}{l}{\textit{Frontier DM}} \\
Jev & $0.022$ & $0.24$ & $0.41$ & $0.12$ & $32.99$ & $9.89$ & $11.00$ & $3.30$ \\
Perplexity & $0.006$ & $0.32$ & $0.11$ & $0.17$ & $8.65$ & $13.43$ & $2.88$ & $4.48$ \\
Mammoth & $0.031$ & $0.29$ & $0.57$ & $0.15$ & $45.88$ & $12.17$ & $15.29$ & $4.06$ \\
D1 Ultra & $0.009$ & $0.27$ & $0.17$ & $0.14$ & $13.40$ & $11.19$ & $4.47$ & $3.73$ \\
\midrule
\multicolumn{9}{l}{\textit{LLM}} \\
Luna LLM & $0.050$ & $2.12$ & $0.94$ & $1.09$ & $25.22$ & $29.41$ & $25.22$ & $29.41$ \\
Sol & $0.871$ & $1.71$ & $16.20$ & $0.88$ & $435.52$ & $23.73$ & $435.52$ & $23.73$ \\
Gemini Argon & $1.169$ & $6.10$ & $21.75$ & $3.15$ & $584.75$ & $84.79$ & $584.75$ & $84.79$ \\
Haiku & $0.139$ & $2.04$ & $2.58$ & $1.05$ & $69.28$ & $28.34$ & $69.28$ & $28.34$ \\
\bottomrule
\end{tabular}
\caption{\textbf{Projected total cost and time of each experiment.}  Occ. mobility refers to the occupational mobility ABM in \citet{del2021occupational}. Average is the mean cost (\$ per 1,000 calls) and time (seconds per call) over the three experiments of Table~\ref{tab:speed_cost}. Each experiment's cost (\$) is its calls times the mean cost per call, and its time (hours) its calls times the mean seconds per call divided by the calls run at once. Purchase Intent, 18,600 calls, 10 at once; Occupational mobility, 500,000 calls (3 for each by a decision model), 10 at once; Rationing, 500,000 calls, 10 at once.}
\label{tab:project_costs}
\end{table}

\end{document}